\documentclass[aps,prb,twocolumn,superscriptaddress,showpacs
]{revtex4-1}	
\usepackage[ngerman,english]{babel} 
\usepackage[utf8]{inputenc}

\usepackage{amsmath,amssymb} 

\usepackage[pdfstartview=FitH,      
            breaklinks=true,        
            bookmarksopen=true,     
            bookmarksnumbered=true  
            ]{hyperref}				

\usepackage{floatrow}    
\newfloatcommand{capbtabbox}{table}[][\FBwidth]

\usepackage{wrapfig}
\usepackage{subfig} 

\usepackage{array}
\usepackage{multirow}

\usepackage{adjustbox} 		
\usepackage{upgreek}		
\usepackage{units}		
\usepackage{mathtools}			
\usepackage{xargs}                      

\usepackage[colorinlistoftodos,prependcaption,textsize=tiny]{todonotes}
\newcommandx{\unsure}[2][1=]{\todo[linecolor=red,backgroundcolor=red!25,bordercolor=red,#1]{#2}}
\newcommandx{\change}[2][1=]{\todo[linecolor=blue,backgroundcolor=blue!25,bordercolor=blue,#1]{#2}}
\newcommandx{\info}[2][1=]{\todo[linecolor=OliveGreen,backgroundcolor=OliveGreen!25,bordercolor=OliveGreen,#1]{#2}}
\newcommandx{\improvement}[2][1=]{\todo[linecolor=Plum,backgroundcolor=Plum!25,bordercolor=Plum,#1]{#2}}
\newcommandx{\thiswillnotshow}[2][1=]{\todo[disable,#1]{#2}}

\newcommand{\etal}{\textit{et al.}}

\usepackage{lipsum}
\usepackage{booktabs}  

\begin{document}


\title{Beyond the gyrotropic motion: dynamic C-state in vortex spin torque  oscillators }


\author{Steffen Wittrock}\email[]{steffen.wittrock@cnrs-thales.fr}\affiliation{Unit\'{e} Mixte de Physique CNRS, Thales, Université Paris-Saclay, 1 Avenue Augustin Fresnel, 91767 Palaiseau, France }
\author{Philippe Talatchian}\altaffiliation[Present address: ]{Institute for Research in Electronics and Applied Physics, University of Maryland, College Park,  20899-6202, MD, USA }\affiliation{Unit\'{e} Mixte de Physique CNRS, Thales, Université Paris-Saclay, 1 Avenue Augustin Fresnel, 91767 Palaiseau, France }
\author{Miguel Romera Rabasa}\altaffiliation[Present address: ]{GFMC, Departamento de Física de Materiales, Universidad Complutense de Madrid, 28040 Madrid, Spain}\affiliation{Unit\'{e} Mixte de Physique CNRS, Thales, Université Paris-Saclay, 1 Avenue Augustin Fresnel, 91767 Palaiseau, France }
\author{Samh Menshawy}\affiliation{Unit\'{e} Mixte de Physique CNRS, Thales, Université Paris-Saclay, 1 Avenue Augustin Fresnel, 91767 Palaiseau, France }
\author{Mafalda Jotta Garcia}\affiliation{Unit\'{e} Mixte de Physique CNRS, Thales, Université Paris-Saclay, 1 Avenue Augustin Fresnel, 91767 Palaiseau, France }
\author{Marie-Claire Cyrille}\affiliation{ Université Grenoble Alpes, CEA-LETI, MINATEC-Campus, 38000 Grenoble, France }
\author{Ricardo Ferreira}\affiliation{ International Iberian Nanotechnology Laboratory (INL), 471531 Braga, Portugal }
\author{Romain Lebrun}\affiliation{Unit\'{e} Mixte de Physique CNRS, Thales, Université Paris-Saclay, 1 Avenue Augustin Fresnel, 91767 Palaiseau, France }
 \author{Paolo Bortolotti}\affiliation{Unit\'{e} Mixte de Physique CNRS, Thales, Université Paris-Saclay, 1 Avenue Augustin Fresnel, 91767 Palaiseau, France }
 \author{Ursula Ebels}\affiliation{ Université Grenoble Alpes, CEA, INAC-SPINTEC, CNRS, SPINTEC, 38000 Grenoble, France }
\author{Julie Grollier}\affiliation{Unit\'{e} Mixte de Physique CNRS, Thales, Université Paris-Saclay, 1 Avenue Augustin Fresnel, 91767 Palaiseau, France }
 \author{Vincent Cros}\email[]{vincent.cros@cnrs-thales.fr}\affiliation{Unit\'{e} Mixte de Physique CNRS, Thales, Université Paris-Saclay, 1 Avenue Augustin Fresnel, 91767 Palaiseau, France }


\date{\today}

\definecolor{light-gray}{rgb}{0.96,0.96,0.96}

\begin{abstract}


In the present study, we investigate a dynamical mode beyond the gyrotropic (G) motion of a magnetic vortex core in a confined magnetic disk of a nano-pillar spin torque nano oscillator. 
It is characterized by the in-plane circular precession associated to a C-shaped magnetization distribution. 
We show a transition between G and C-state mode which is found to be purely stochastic in a current-controllable range. 
Supporting our experimental findings with micromagnetic simulations, we believe that the results provide novel opportunities for the dynamic and stochastic control of STOs, which could be interesting to be implemented for example in neuromorphic networks.

\end{abstract}


\maketitle


\section{Introduction}

Spin torque nano oscillators (STNOs) have attracted considerable attention as next-generation multifunctional spintronic devices \cite{Locatelli2013,Ebels2017} over the last decade. 
Potential applications reach from high data transfer rate hard disk reading \cite{Sato2012}  and wide-band high-frequency communication\cite{Choi2014,Ruiz-Calaforra2017,Kreissig2017,Litvinenko2019} 
to broadband microwave energy harvesting \cite{Fang2019} or frequency detection \cite{Jenkins2016,Louis2017}. Furthermore, their exploitation in the reconstruction of bio-inspired networks for neuromorphic computing \cite{Torrejon2017,Romera2018} creates new opportunities for novel unconventional computing architectures. 
After more than a decade of research, among different realizations of STNOs, vortex based STNOs (STVOs) have been considered a model system for nonlinear dynamics\cite{Pribiag2007,Dussaux2010} exhibiting the best performances in terms of coherence and tunability. 
High output emission powers are simultaneously reached by the large magnetoresistive ratio of magnetic tunnel junctions (MTJs).  
While different dynamic characteristics of the non-uniform vortex magnetization distribution have been extensively studied -- including specifically higher order modes corresponding to radial or orthoradial spin wave modes in the vortex tail \cite{Novosad2002,Guslienko2005,Ivanov2005,Taurel2016} --, the exploitation in STVOs is most commonly limited to the vortex core gyrotropic (G) motion \cite{Thiele1973}.
Indeed, only few studies report physics invoking the motion of the vortex core beyond this fundamental mode in STVOs. 
For instance, Strachan \etal  \cite{Strachan2008} report on the current-induced switching process of a uniform magnetization direction mediated by a vortex C- and G-state. 
In Refs. \cite{Jin2009,Aranda2010}, the switching of a vortex core's polarity through a transient C-state mode is described, however, at pillar diameters of $d \lesssim 120\,$nm (depending on the material's exchange length) and an out-of-plane magnetized polarizer \cite{Jin2009,Aranda2010,Kawada2014}.  
Under the same geometric limitations, an excited S-state \cite{Aranda2010} and C-state \cite{Kawada2014} of weakly \cite{Aranda2010} or significantly  \cite{Kawada2014}  increased frequency compared to the G-state is found in Refs. \cite{Aranda2010,Kawada2014}. 
In this work, we demonstrate the characteristics of a spin transfer induced dynamic C-state with slightly lower frequency than the G-state, but likewise with $df/dI> 0$ and large coherence and emission power in a MTJ based STNO with an in-plane polarizer. 
Both in experiment and micromagnetic simulation, we investigate  the stochastic nature of the transition between the two modes, facilitating potential random features\cite{Jenkins2019} that could be also exploited in the emerging field of neuromorphic spintronics.   

\section{Experiments}
\label{sec:measurements}


The studied vortex STNO devices are circular shaped MTJs of $300\,$nm dot diameter, in-plane polarized reference layer and a $7\,$nm NiFe free layer (see supplementary\cite{supplementary_C-state_2020}).

\begin{figure}[tbh!]
  \centering  
  \captionsetup[subfigure]{}
  \subfloat[\label{fig:merged_simulation-exp}]{
 \includegraphics[width=0.62\columnwidth]{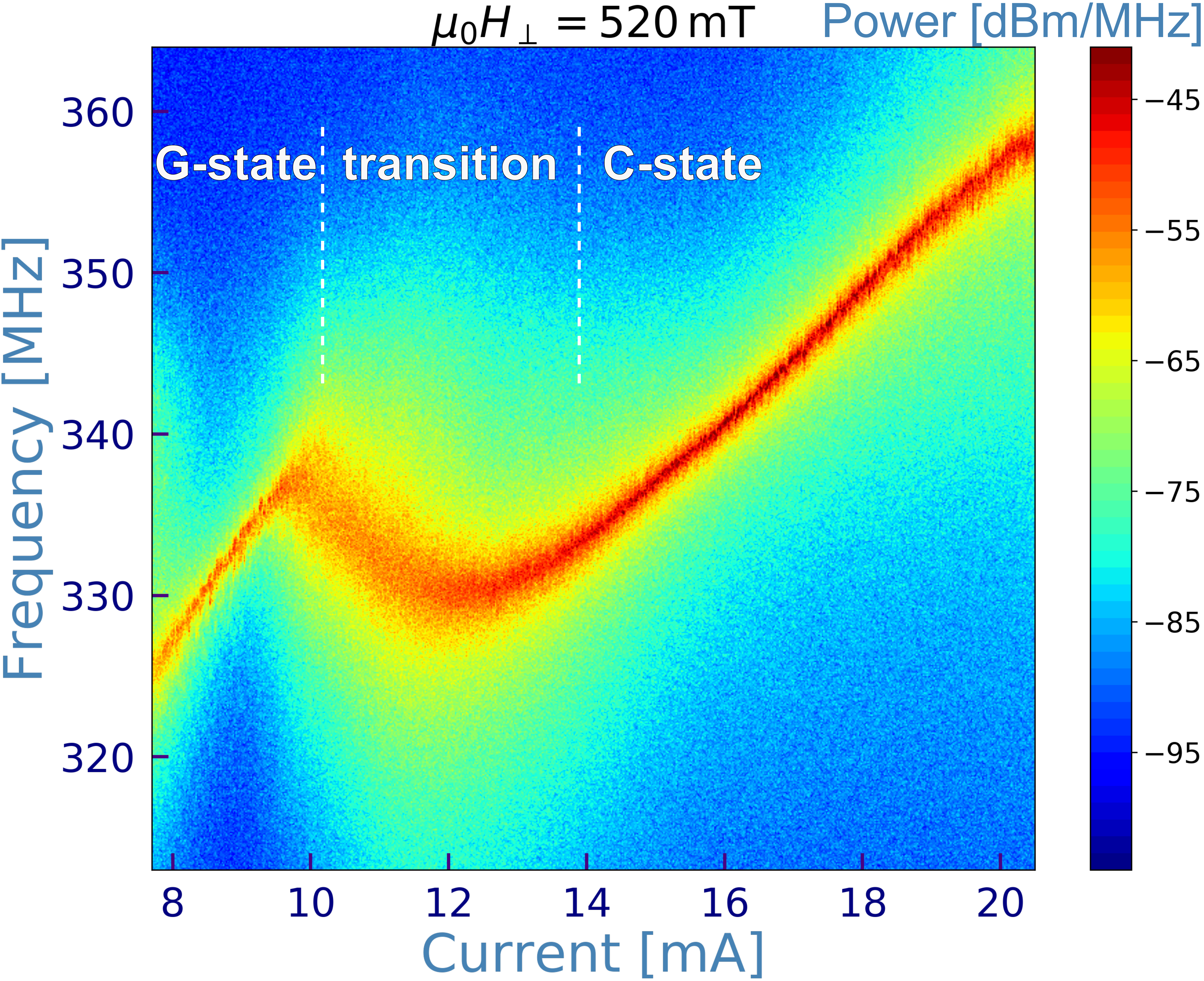}
}
\\
\subfloat[\label{fig:basic_properties}]
{
\includegraphics[width=0.66\textwidth]{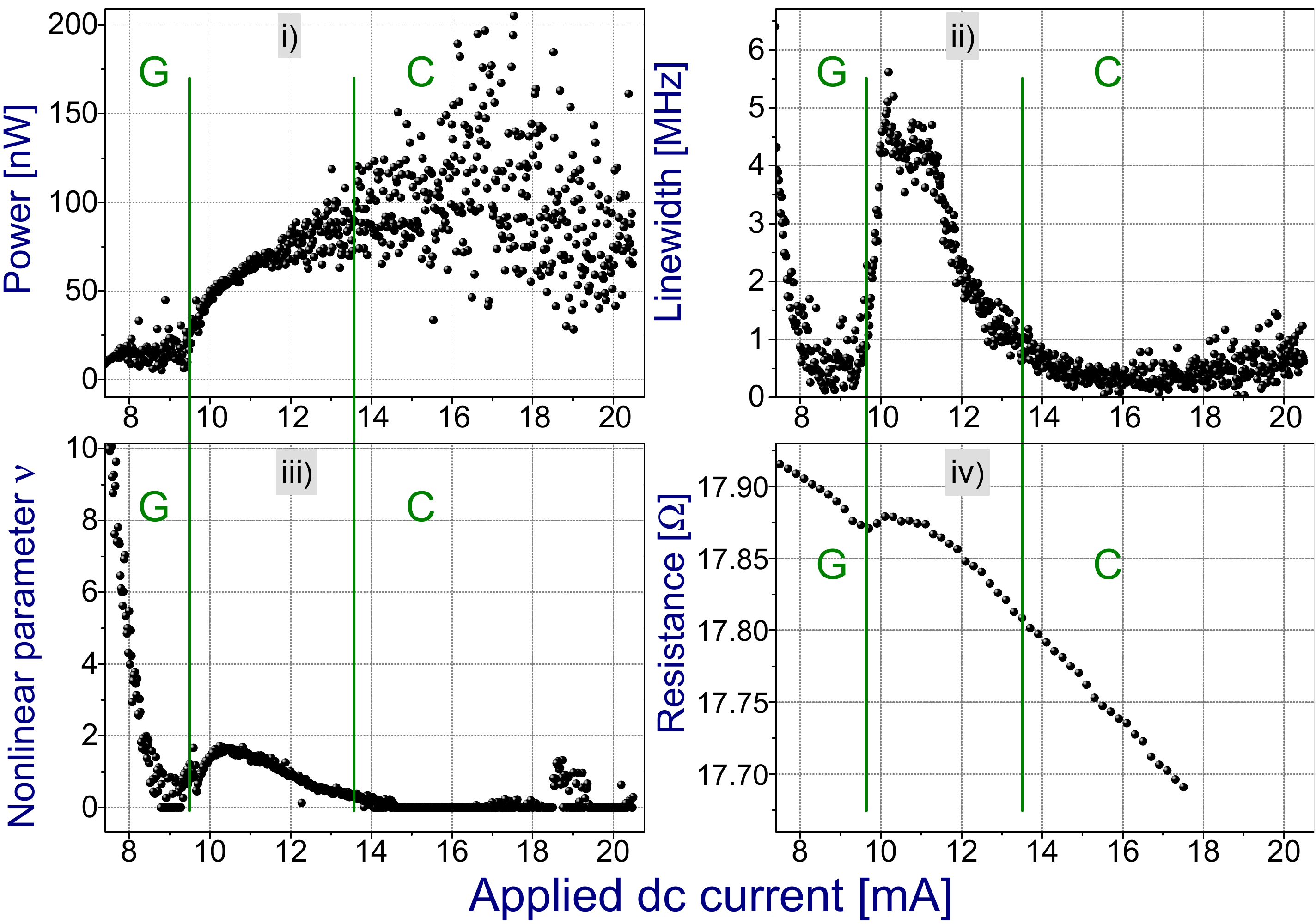} 
}
  \caption[]{Measurement of the dynamic states.  (a) Power emission spectra  and (b) corresponding extracted properties of the oscillation vs. $I_{dc}$: (b.i) emission power, (b.ii) linewidth FWHM, (b.iii) nonlinear parameter $\nu$, and (b.iv) average sample resistance $R_0$.  An out-of-plane  magnetic field of $520\,$mT was applied.  } 
  \label{fig:exp-data_1}
\end{figure}

In fig. \ref{fig:exp-data_1}, we show typical measurements presenting different dynamic states: 
In fig. \ref{fig:merged_simulation-exp}, the power emission spectra of the current supplied STVO are plotted. 
For $I_{dc} \in [7;9.7]\,$mA, a quasi-linear evolution of the oscillation frequency with the current is observed. 
As shown in fig. \ref{fig:basic_properties},  the oscillation parameters in this interval evolve as expected for the gyrotropic vortex motion: The linewidth decreases with the injected dc current $I_{dc}$  and the oscillation amplitude increases, as also the vortex core position radius on its stable limit cycle grows with $I_{dc}$. 
In the following, we refer to these characteristics in the mentioned interval as the gyrotropic (G) state. 

For $I_{dc} \in [9.7;13.5]\,$mA, the frequency decreases rapidly with strongly enhanced linewidth (fig. \ref{fig:basic_properties}.ii) and nonlinearity parameter\cite{Slavin2009,Wittrock2019_PRB} $\nu$ (fig. \ref{fig:basic_properties}.iii)   compared to the low-linewidth G-state.  
This regime represents a transition between two  frequency values, whereas the frequency at the higher current value is lower. 
It is generally observed and the transition can be more (see supplementary) or less (fig. \ref{fig:exp-data_1}) discrete at different values of $I_{dc}$ depending on the chosen experimental parameters. 
Interestingly, the average sample resistance, which usually decreases monotonically with $I_{dc}$ due to the TMR bias dependance, exhibits a relative increase within the transition (fig. \ref{fig:basic_properties}.iv) indicating a more antiparallel magnetization configuration. 
A third dynamical regime appears for $I_{dc} \gtrsim 13.5\,$mA, in which another quasi-linear frequency evolution with the current is observed. 
The linewidth FWHM stabilizes after the transition down to  $\sim 300\,$kHz, a similar value as obtained as well in the G-state. 
Importantly, the emission power is much higher than for the G-state oscillation (factor $3$-$4$ in the presented measurement) and does furthermore  not systematically change with the applied dc current. 
Note that however, the power is subject to rather large fluctuations at constant linewidth, potentially caused by an enhanced flicker noise in this regime\cite{Wittrock2019_PRB,Wittrock2020-SciRep}. 
For very high current values above $17\,$mA, the linewidth increases again.  
As for the nonlinearity parameter $\nu$, it first reduces almost to zero after the transition and consequently, is practically irrelevant in the measured C-state regime. However, measurements indicate that it significantly increases again for even higher currents in the C-state.

 \begin{figure}[!hbt]
 	\centering
 	{\includegraphics[width=0.658\textwidth]{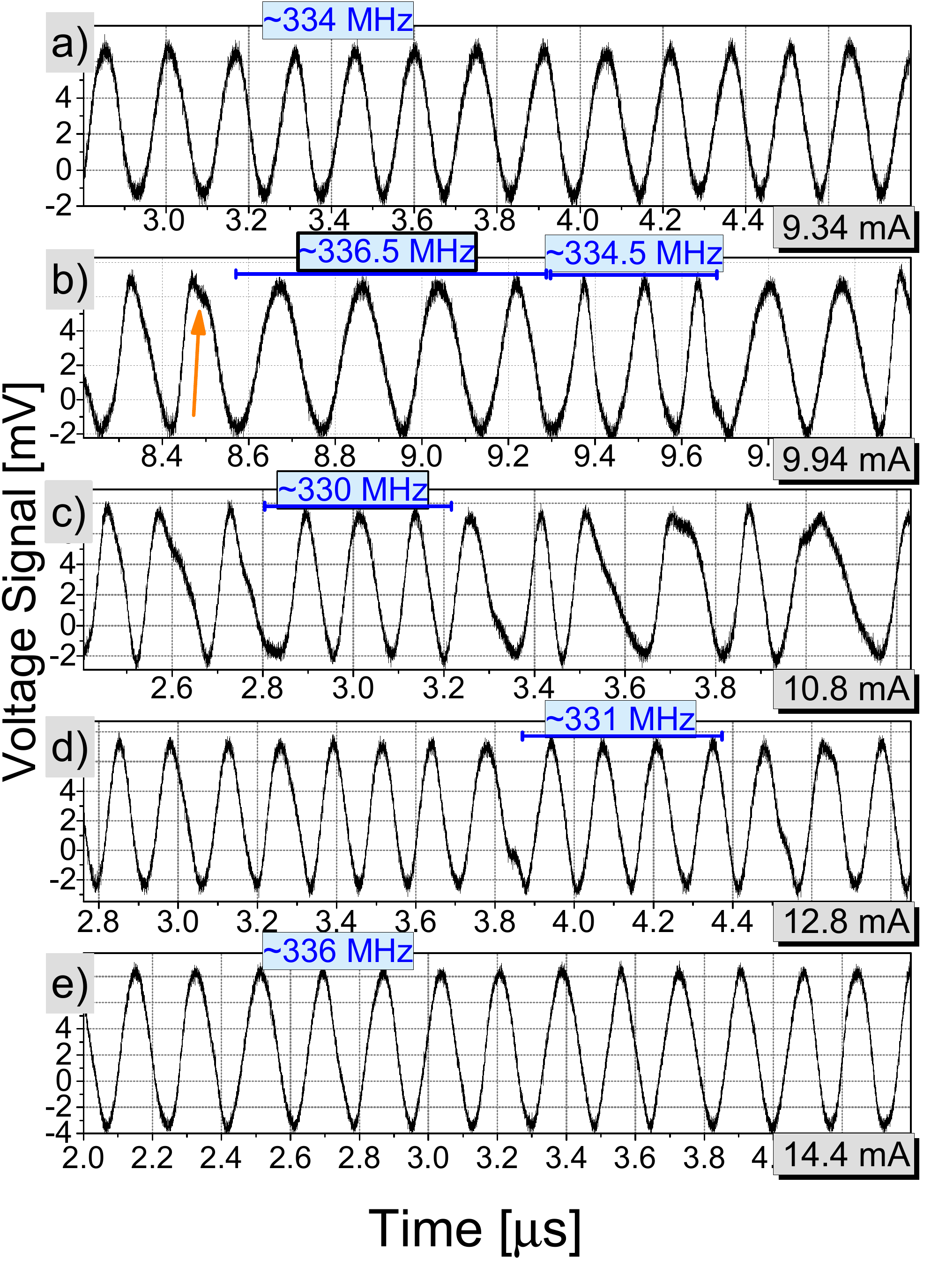}  } 
 	\caption{Measured heterodyne voltage signal vs. time at different current values in the (a) G-state, (b-d) transition regime and (e) C-state. The given frequencies (blue) are average values over the corresponding marked (b-d) or total (a,e) sequence.  }
 	\label{fig:time-traces}
 \end{figure}

In fig. \ref{fig:time-traces}, we record the oscillation time signal using a heterodyne detection technique (signal down-conversion to $5$-$10\,$MHz via high-side injection and low-pass filtering) in order to resolve  small variations of frequency. 
Note that then, one period of the heterodyne signal roughly corresponds to $\sim 50$ periods of the pure signal. 
At the beginning of the transition regime at $9.94\,$mA  (fig. \ref{fig:time-traces}.b),  the oscillator is mainly in the  G mode ($336.5\,$MHz in the intervall $t\sim [8.5;9.3]\,\upmu$s).  
In addition, for example at $t\approx 8.5\,\upmu$s, the signal shows a kink (orange arrow), which is interpreted as a part of the oscillation in the C-state or potentially pinning at the disk boundary for a short moment of time. 
  For a short time in the interval $\sim [9.3;9.7]\,\upmu$s, the system oscillates at a slightly different frequency: 
  We assume here, that the oscillator sometimes performs even several oscillation periods in the C-state before switching back to the G-state (see also the supplementary material\cite{supplementary_C-state_2020}). 
For a larger dc current $I_{dc} = 10.8\,$mA in the center of the transition regime (fig. \ref{fig:time-traces}.c), the time signal is much less coherent, which we interpret as a  regular switching between the states together with potential pinning events. 
For $I_{dc}=12.8\,$mA (fig. \ref{fig:time-traces}.d), at the end of the transition, the vortex core is again only rarely constrained at the potential barrier. 
For currents well above or below the transition region, and hence for the G- or C-state (fig. \ref{fig:time-traces}.a and \ref{fig:time-traces}.e, resp.), the time traces show a stable oscillation signal without observed instabilities and state switching events.

\section{Micromagnetic simulation}
\label{sec:simulation}


\begin{figure*}[bth!]
\centering
\floatsetup[subfigure]{}
\ffigbox[\textwidth]{
\begin{subfloatrow}[2]
\thisfloatsetup{capbesidesep=none,objectset=centering,%
capbesideframe=yes,capbesideposition={right,bottom}}
\ffigbox[\FBwidth]{
\includegraphics[height=0.344\textwidth]{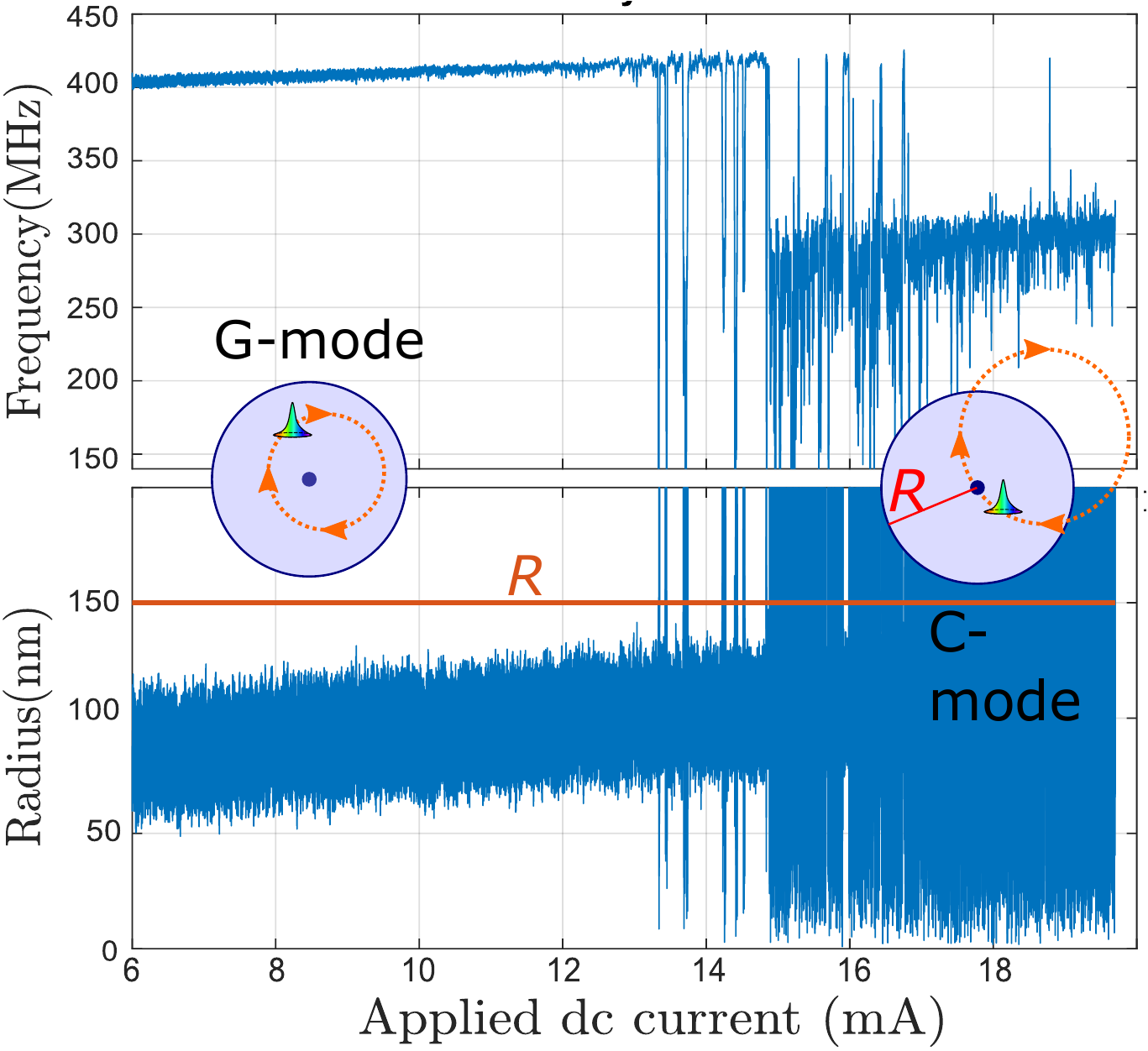}}{ \caption{ }  \label{fig_C-state:sim_T=300_FLT=0,4_Frequ_radius}}
\thisfloatsetup{capbesidesep=none,objectset=centering,%
capbesideframe=yes,capbesideposition={right,bottom}}
\ffigbox[\FBwidth]{
\includegraphics[height=0.352\textwidth]{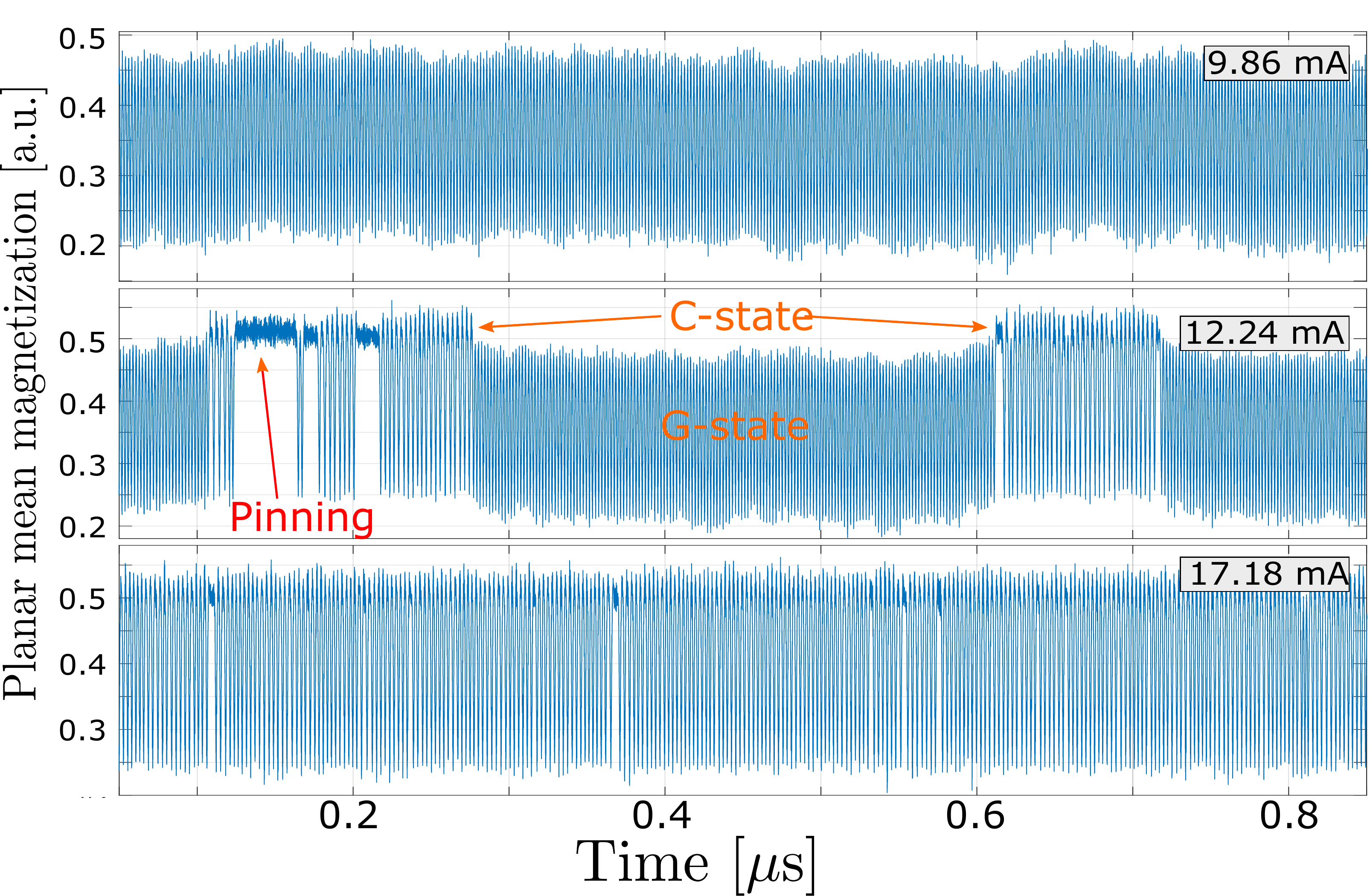}}{ \caption{ } \label{fig_C-state:sim_T=300_FLT=0,4_TimeTraces}}
\end{subfloatrow}

\vspace{2ex}    
    
   \thisfloatsetup{capbesidesep=columnsep,objectset=centering,%
capbesideframe=yes,capbesideposition={left,center}}
    \fcapside[0.5\textwidth]{
     \begin{subfloatrow}[2]
      \ffigbox[\FBwidth]{%
    \includegraphics[width=0.92\columnwidth]{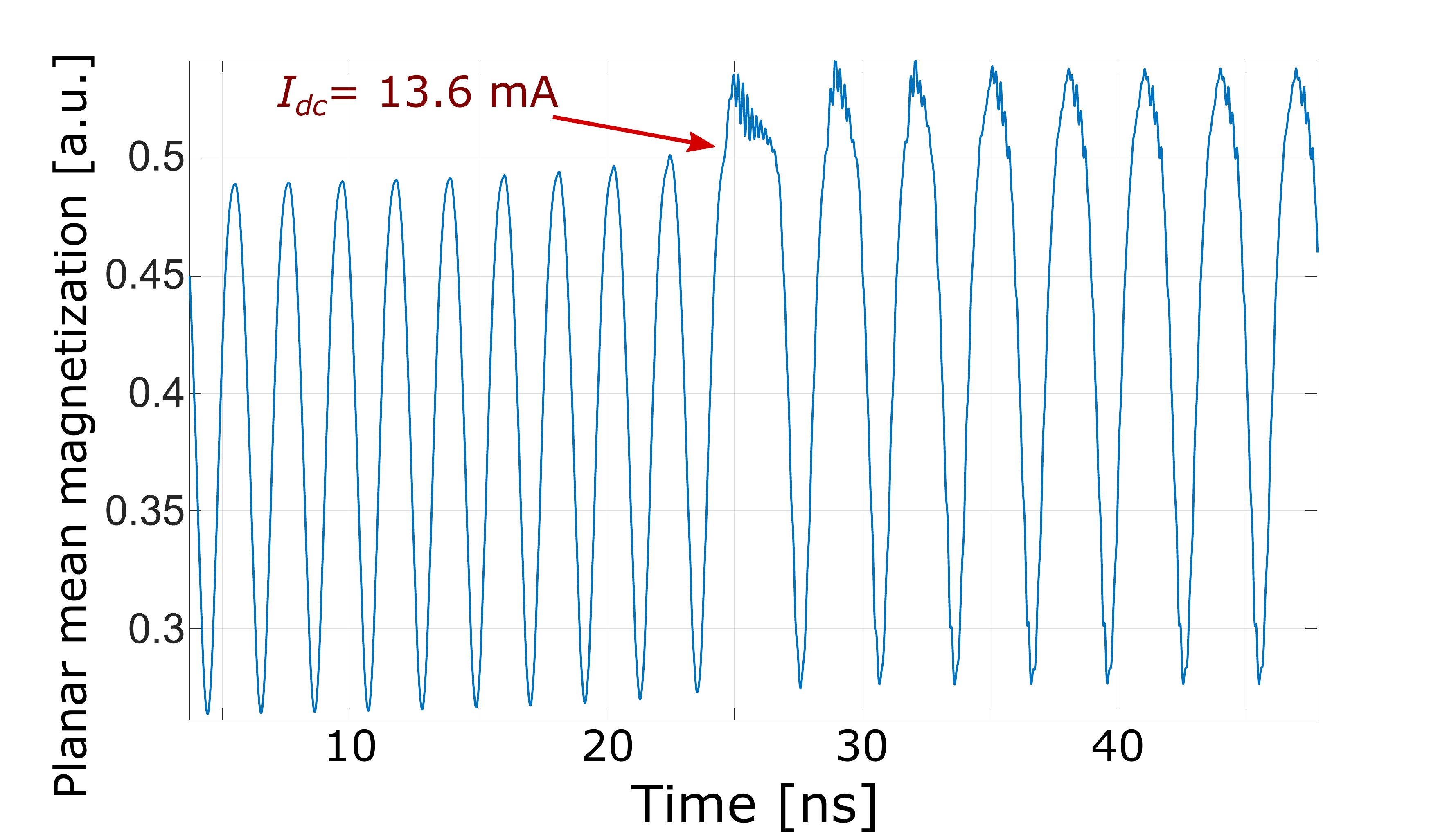} }{\caption{} 
    \label{fig:T=0_with_FLT}  }  
     \end{subfloatrow}
     }
          { \caption[Parameter space of the PLL+STNO phase noise PSD ]{Micromagnetic simulation of the magnetization dynamics  
  of gyrotropic (G) and C-state and their transition. {(a)} Frequency  and vortex core radius evolution with applied dc current. {(b)} Planar mean  magnetization vs. time at different current values.  (c) Discrete transition when slightly increasing $I_{dc}$ with time at $T=0\,$K. 
}
 \label{fig_C-state:sim_T=300_FLT=0,4} }
    }
 {}
\end{figure*}

In order to further investigate the nature of the different dynamic modes, we perform mumax$^3$ micromagnetic simulations \cite{mumax} of the magnetization dynamics  (see supplementary\cite{supplementary_C-state_2020}). 
For this purpose, we include the different spin-torque contributions that can control the current driven magnetization dynamics, namely the Oersted field,  Slonczewski (STT) and the field-like torque (FLT)\cite{Ralph}.

Simulation results are shown in fig. \ref{fig_C-state:sim_T=300_FLT=0,4} and   reproduce the frequency characteristics of the measurements as described for the experimental data:
 Up to  $I_{dc} \sim 13\,$mA, we find  in fig. \ref{fig_C-state:sim_T=300_FLT=0,4_Frequ_radius} the gyrotropic mode, i.e. the quasi-circular motion of the vortex core takes place inside the nanodisk as depicted in fig. \ref{fig_C-state:sim_T=300_FLT=0,4_Frequ_radius}'s inset (see also videos in the supplementary\cite{supplementary_C-state_2020}). 
 Note that the vortex gyration is not symmetric relative to the disk center due to the FLT\cite{Dussaux2012}, whose efficiency compared to the STT is set to $\xi_{fl} = 0.4$, a reasonable value for MTJ STVOs\cite{Chanthbouala2011}. 
For $I_{dc} \gtrsim 13\,$mA, we observe a transition to the C-state which stabilizes for higher dc currents. 
At this point, the introduced expression \textit{"C-state"}\ also becomes clear: The simulations reveal that at higher currents $I_{dc} \gtrsim 17\,$mA, the vortex core is not continuously found to be inside the  nanodisk. 
Instead,  it establishes that for a part of the oscillation period, the vortex core is inside and for the other part, it is imaginarily outside the disk, i.e. in the latter the in-plane magnetization of the vortex tail forms a \textit{C} which precesses following an imaginary vortex core. 
The situation is depicted in fig. \ref{fig_C-state:sim_T=300_FLT=0,4_Frequ_radius}'s inset. 
Furthermore, the lower panel of fig. \ref{fig_C-state:sim_T=300_FLT=0,4_Frequ_radius} clearly proves  the vortex core  oscillation radius partly inside, partly outside the nanodisk of radius $R$.  
In the supplementary material\cite{supplementary_C-state_2020}, movies of the occurring dynamic situations can be found. 
We also perform simulations without FLT, i.e.  $\xi_{fl} = 0$  (see supplementary material\cite{supplementary_C-state_2020}). 
The simulation results in this case are fundamentally different and describe after the transition from the G-state  a pure C-state (no renucleation of the vortex core) with significantly higher frequency in the order of $1\,$GHz. 
 These results qualitatively reproduce what has been obtained by Kawada \etal\cite{Kawada2014} for the experimental case of a GMR based STVO, where indeed the  FLT is negligible. 
Thus, the FLT in MTJs is an important parameter in order to control the described C-state characteristics.

In the transition between the two modes in fig. \ref{fig_C-state:sim_T=300_FLT=0,4_Frequ_radius}, different situations can occur: 
Firstly,  the system incoherently switches between both states, performing several periods in the G- and several periods in the C-state, as also observed experimentally. 
Secondly, the vortex core pins at the disk boundary for a certain period of time, hence stopping oscillator operation until it resumes dynamics. 
In fig. \ref{fig_C-state:sim_T=300_FLT=0,4_TimeTraces}, exactly this behavior can be observed through the planar mean magnetization: For  $I_{dc}=9.86\,$mA, the system steadily oscillates in the G-state. 
In this case, oscillations of the in-plane mean magnetization can be obtained because 
of the asymmetry in the gyrotropic motion (not centered) induced by the FLT,  particularly included  in the simulations.  
For $I_{dc}=12.24\,$mA, we observe a stochastic  switching between the states: For example, from $0.6< t < 0.72\,\upmu$s, the system is in the C-state. 
It has a lower frequency and the planar magnetization, due to the lack of the vortex core, is slightly enhanced. 
From $0.1< t < 0.2\,\upmu$s, it can be observed that the vortex core pins at the disk boundary resulting in a quasi-constant in-plane magnetization until readopting oscillations. 
Comparing to experimental measurements (fig. \ref{fig:exp-data_1}), the switching between the two states translates into a broadening of the effective linewidth. 
Note that there, the likelihood of vortex core pinning at the boundary and thus, the range of the transition depends on several experimental parameters, such as the applied magnetic field, sample dimensionality, free layer material, etc. resulting in different characteristics of the transition (shown in the supplementary\cite{supplementary_C-state_2020}). 
For $I_{dc}=17.18\,$mA in fig. \ref{fig_C-state:sim_T=300_FLT=0,4_TimeTraces}, the system oscillates in the C-state providing coherent oscillations, although the vortex core pins at the boundary at some very short moments from time to time. The number of pinning events decreases at even higher currents. 
Notably, figs. \ref{fig_C-state:sim_T=300_FLT=0,4_Frequ_radius} \& \ref{fig_C-state:sim_T=300_FLT=0,4_TimeTraces} show that the transition region between G- and C-state is completely stochastic. The state probability can be continuously tuned within the transition until, at higher current values, the C-state is stabilized, and at lower current values, the G-state.

In fig. \ref{fig:T=0_with_FLT}, we show the transition between G- and C-state when performing the simulation without thermal noise at $T=0\,$K. 
When slightly increasing the current to the critical transition current at $\sim 24\,$ns, the transition from G- to C-state oscillations is observed. 
Along with the reduced frequency in the C-state, also an asymmetry within one period (equally present for $T=300\,$K) is recognized: When the vortex is expelled (high mean in-plane magnetization) for approximately half a period, the frequency is slightly reduced compared to the other half-period when the vortex core is located inside the nanodisk. 
Moreover, the transition between G- and C-state is discrete and both situations exhibit stable oscillations, indicating that the unstable transition region and stochastic state switching described beforehand  is noise induced due to thermal fluctuations.

\section{Discussion}  


The emission power in the C-state is considerably higher than in the G-state (fig. \ref{fig:basic_properties}). 
Comparing with the performed simulations, we attribute this to a  larger active magnetic volume contributing to the magnetoresistive rf signal. 
The quasi-zero nonlinearity parameter $\nu$ in the C-state might be due to a changed confinement when the magnetization distribution is not vortex-like. 
Note that $\nu$ might again become $\nu > 0$ at even higher applied current values. 
Moreover, the nonlinearity parameter within the transition region is rather an effective value as the system is bistable and the single-mode hypothesis does not hold any more. 
However, as the frequency tunability linked to the large nonlinear parameter $\nu$ is strongly enhanced in this regime, we anticipate that the locking ability to an external stimulus is significantly improved here. 
Especially interesting from an application point of view is the vicinity of a high $\nu$ to a $\nu\approx 0$ regime: With the locking efficiency proportional to $\sqrt{1+\nu^2}$, the ability to respond and synchronize to an external signal could thus be easily tuned by only changing the applied dc current.


 

\section{Conclusion}


In this study, we observe and describe a dynamic C-state beyond the vortex gyrotropic (G) motion in experiment and, consistently, in micromagnetic simulations. 
Induced by the applied dc current, 
the observed G- to C-state transition is found to be facilitated by the field-like torque in MTJ based STVOs and notably completely stochastic in the presence of thermal noise.  
The state probability can be continuously tuned within the transition regime until, at higher current values, the C-state is stabilized, and at lower current values, the G-state. 
We believe that the stochastic feature might be exploited for cognitive computing,
for which STNOs have recently been implemented in numerous ways. Besides oscillations,
stochasticity is an important feature for energy efficient neuromorphic operations\cite{Grollier2016}. 
The combination of
both, nonlinear oscillations and stochasticity, and furthermore the ability to synchronize or respond to external stimuli, in one unique spintronic device can open the path to
more multifunctional neuron-like building blocks necessary for neuromorphic computing\cite{Grollier2016}.

\section*{\footnotesize{Supplementary Material}}

See supplementary material\cite{supplementary_C-state_2020} for details on the samples and  a further experimental characterization of the occurring transition from G- to C-state. This includes the demonstration of 
a more discrete transition and its characterization with the applied magnetic field. Furthermore, simulation results for $\text{FLT}=0$ are shown and videos of the different regimes included.

\section*{\footnotesize{Acknowledgments}}

S.W. acknowledges financial support from Labex FIRST-TF under contract number ANR-10-LABX-48-01. 
M.R. acknowledges support from Spanish MICINN (PGC2018-099422-A-I00) and Comunidad de Madrid (Atracción de Talento Grant No. 2018-T1/IND-11935). 
The work is supported by the French ANR project ''SPINNET'' ANR-18-CE24-0012.

\section*{\footnotesize{Data Availability Statement}}

The data that support the findings of this study are available from the corresponding author upon reasonable request.
\bibliography{literatur_promo}

\end{document}


\flushbottom
\maketitle

\thispagestyle{empty}


In this supplementary information, we first comment further on the performed experiments. 
After a more detailed description of the measured sample system, we present captured data of the gyrotropic (G) and the C-state, complementing the data already shown in the main article. 
Moreover, we also characterize the measured evolution of the transition regime between the states with the applied magnetic field. \\
Regarding the simulations, we subsequently give additional simulational details and furthermore present simulation results at zero field-like torque $\text{FLT}=0$. 
Apart from this document, \textbf{videos} of the simulation of magnetization dynamics for the different cases can be found in the supplementary material.

\section{Measurement}

\subsection{Sample details}

The experimental measurements presented in this work have been performed on vortex based spin transfer oscillators (STVOs). This type of STNOs exploits the dynamics of a magnetic vortex in a circular shaped nanodisk and converts it into an electric rf signal due to the magnetoresistive effect\cite{Dussaux2010}.

The studied samples consist of a pinned layer made of a conventional synthetic antiferromagnetic stack (SAF), a MgO tunnel barrier and a NiFe free layer in a magnetic vortex configuration.
The  magnetoresistive ratio related to the tunnel magnetoresistance effect (TMR) lies around $50\,$\% at room temperature.  
The SAF is composed of \selectlanguage{ngerman} PtMn($20$)/""Co$_{70}$Fe$_{30}$($2$)/""Ru($0.85$)/""Co$_{40}$Fe$_{40}$B$_{20}$($2.2$)/""Co$_{70}$Fe$_{30}$($0.5$) and the total layer stack is Ta(3)/""CuN(30)/""Ta(5)/""SAF/""MgO($1$)/""Co$_{40}$Fe$_{40}$B$_{20}$($1.5$)/""Ta($0.2$)/""Ni$_{80}$Fe$_{20}$(7)/""Ta(5)/Ru(7), with the nanometer layer thickness in brackets.
\selectlanguage{english}
%
%
%
The layers are deposited on high resistivity SiO$_2$ substrates by ultrahigh vacuum magnetron sputtering, subsequently annealed for $2\,$h at $T=330\,^{\circ}$C at an applied magnetic field of $1\,$T along the SAF's easy axis, and patterned using e-beam lithography and Ar ion etching. 
The $0.2\,$nm Ta-layer decouples the growth of CoFeB and NiFe, utilizing the high tunnel magnetoresistance (TMR) ratio of the crystalline CoFeB-junction and the magnetically softer NiFe for the vortex dynamics. 
The circular tunnel junctions have an actual diameter of $2R = 300\,$nm.

\subsection{Transition characteristics}

\begin{figure}[bth!]
  \centering  
  \subfloat[  \label{fig_C-state:spectra_and_spectrum_685mT_discrete_transition-1}]
  {  
  \includegraphics[width=0.415\textwidth]{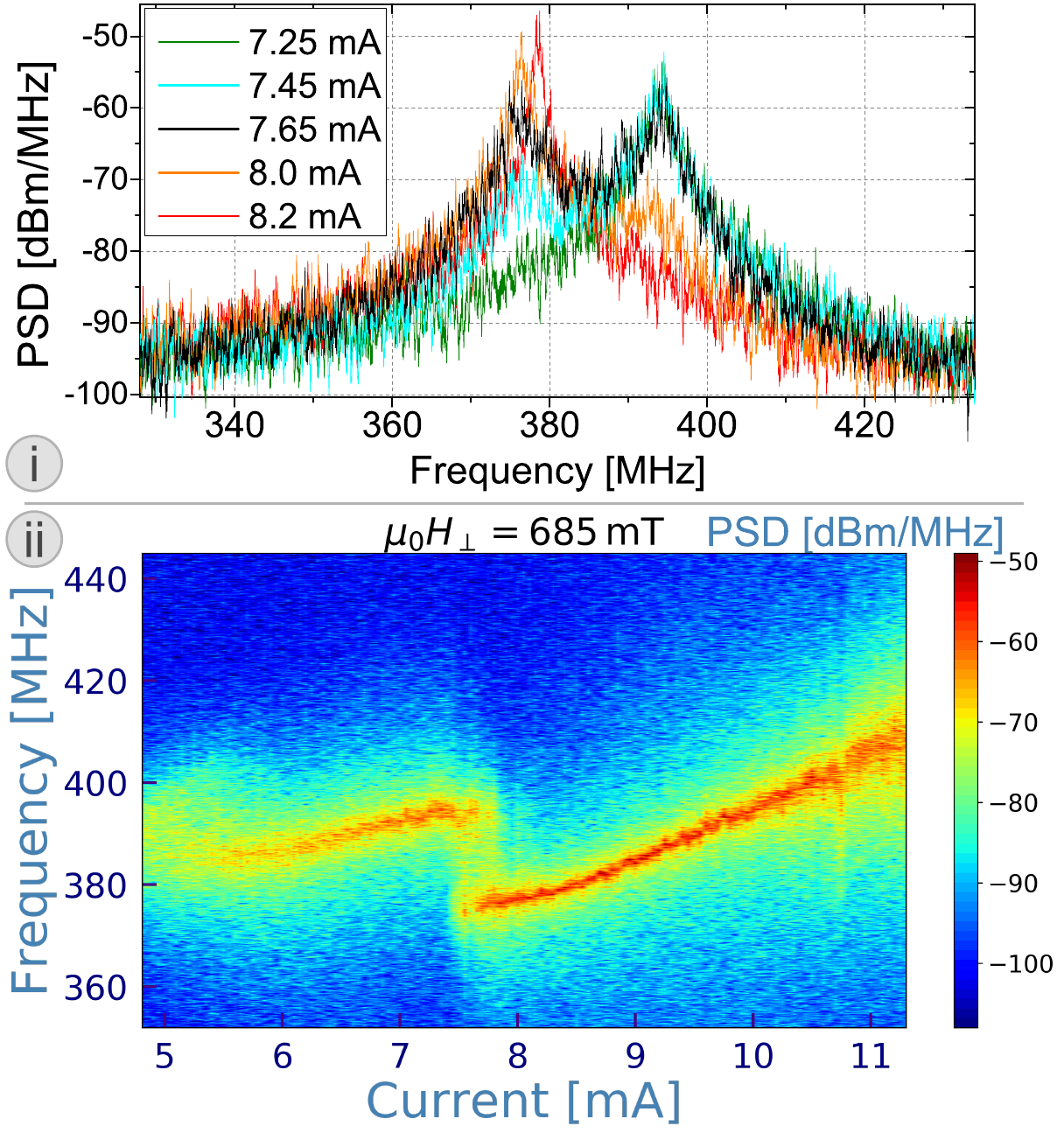}  }   
  \hfill 
 \subfloat[  \label{fig_C-state:spectra_experiment_RTN-2}] 
  { 
  \includegraphics[width=0.4\columnwidth]{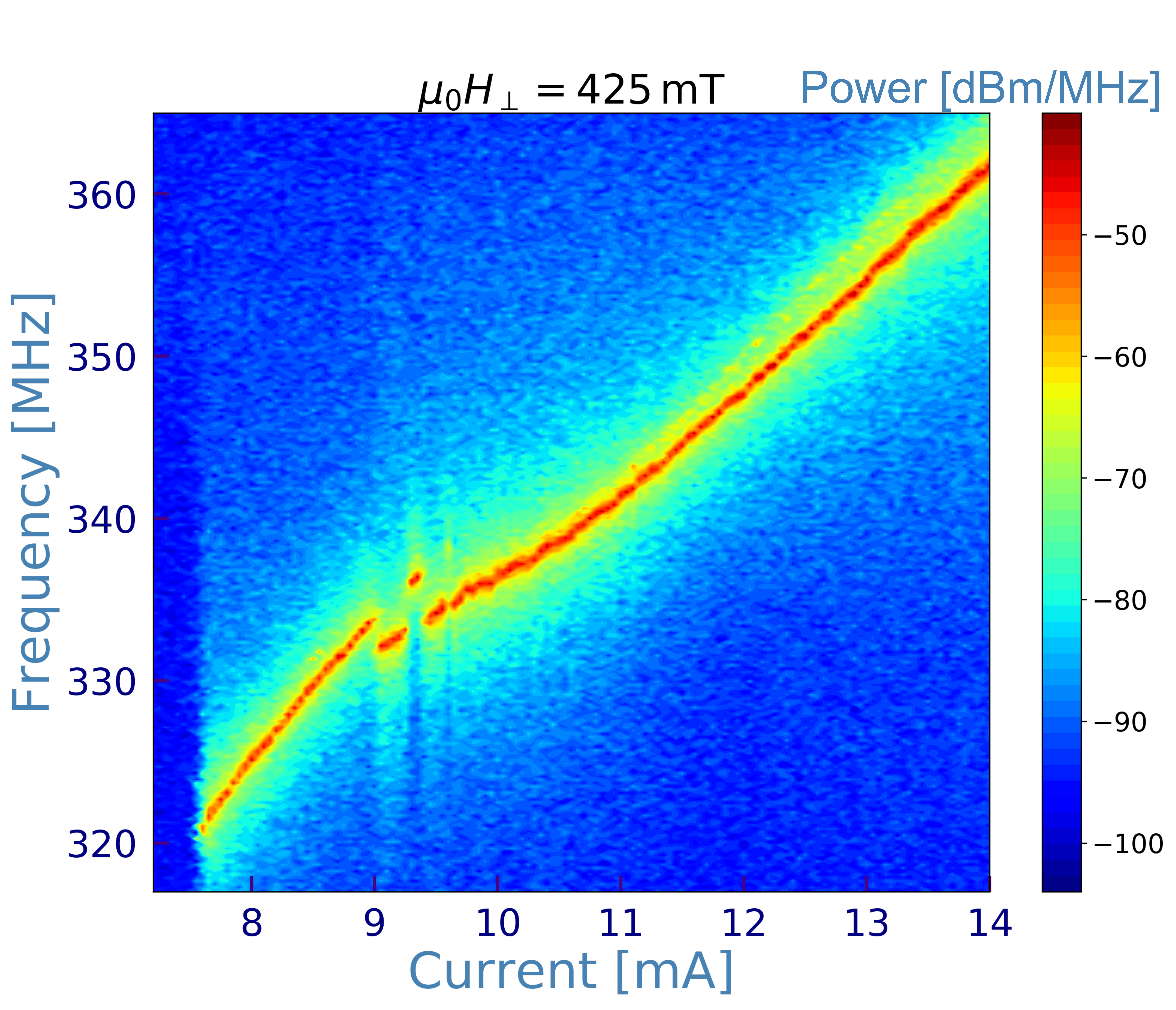}   } 
  \caption[Frequency spectra and transition vs. $I_{dc}$ in different measurements]{Frequency spectra vs. $I_{dc}$ in different measurements (a) and (b) showing a G$\rightarrow$C-state transition.  In (a.i) the frequency spectra for different current values in or close to the transition, corresponding to the frequency spectra surface plot in (a.ii), are shown. 
}
  \label{fig_C-state:other-spectra}
\end{figure}

The transition between the modes in fig. 2 of the main paper is relatively smooth. 
In fig. \ref{fig_C-state:other-spectra}, we show measurements carried out on different spin torque vortex oscillator (STVO) devices under different experimental conditions (same wafer, but different STVO). 
This second measurement series is shown because a more discrete transition between the two states (fig. \ref{fig_C-state:spectra_and_spectrum_685mT_discrete_transition-1}) or some stochastic switching (fig. \ref{fig_C-state:spectra_experiment_RTN-2}) is observed in the spectra, generally highlighting that the transition characteristics can be changed through the experimental parameters. 
Indeed, the measured spectrum in the transition, shown in fig. \ref{fig_C-state:spectra_and_spectrum_685mT_discrete_transition-1}.i, even shows two distinct frequency peaks becoming more or less pronounced depending on the position in the transition regime. 
In fig. \ref{fig_C-state:spectra_experiment_RTN-2}, the transition at $I_{dc} \in \sim [9;11.5]\,$mA is also rather discrete and the system furthermore  switches between the two states when sweeping $I_{dc}$ inside the transition. 
The shown characteristics indicate that the transition from G- to C-state is stochastic.

\begin{figure}[bth!]
  \centering  
 \subfloat[  \label{fig_C-state:TimeTraces-b}] 
  {  \raisebox{0.18cm}{ \hspace{0.0cm} 
  \includegraphics[width=0.4\textwidth]{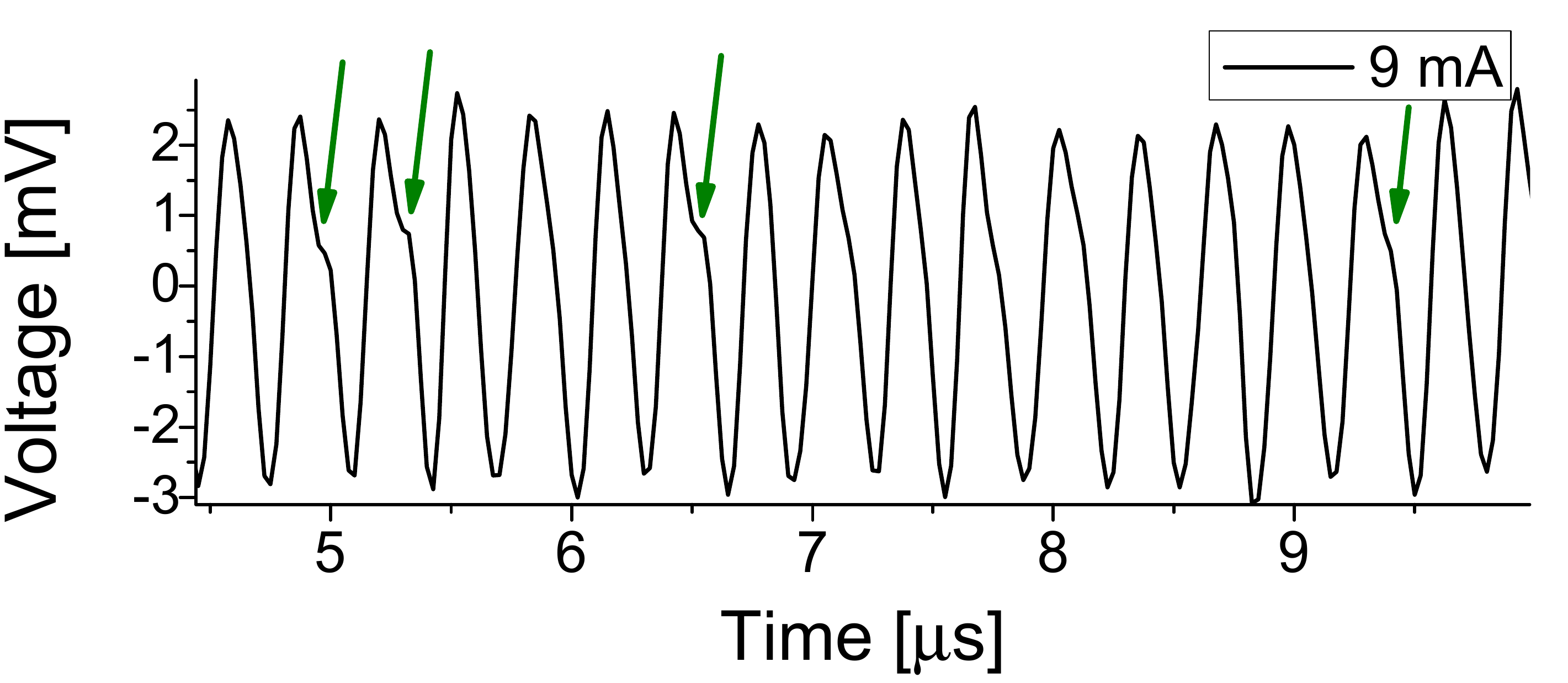}   } } 
   \subfloat[  \label{fig_C-state:TimeTraces-c}] 
  {  
  \includegraphics[width=0.545\textwidth]{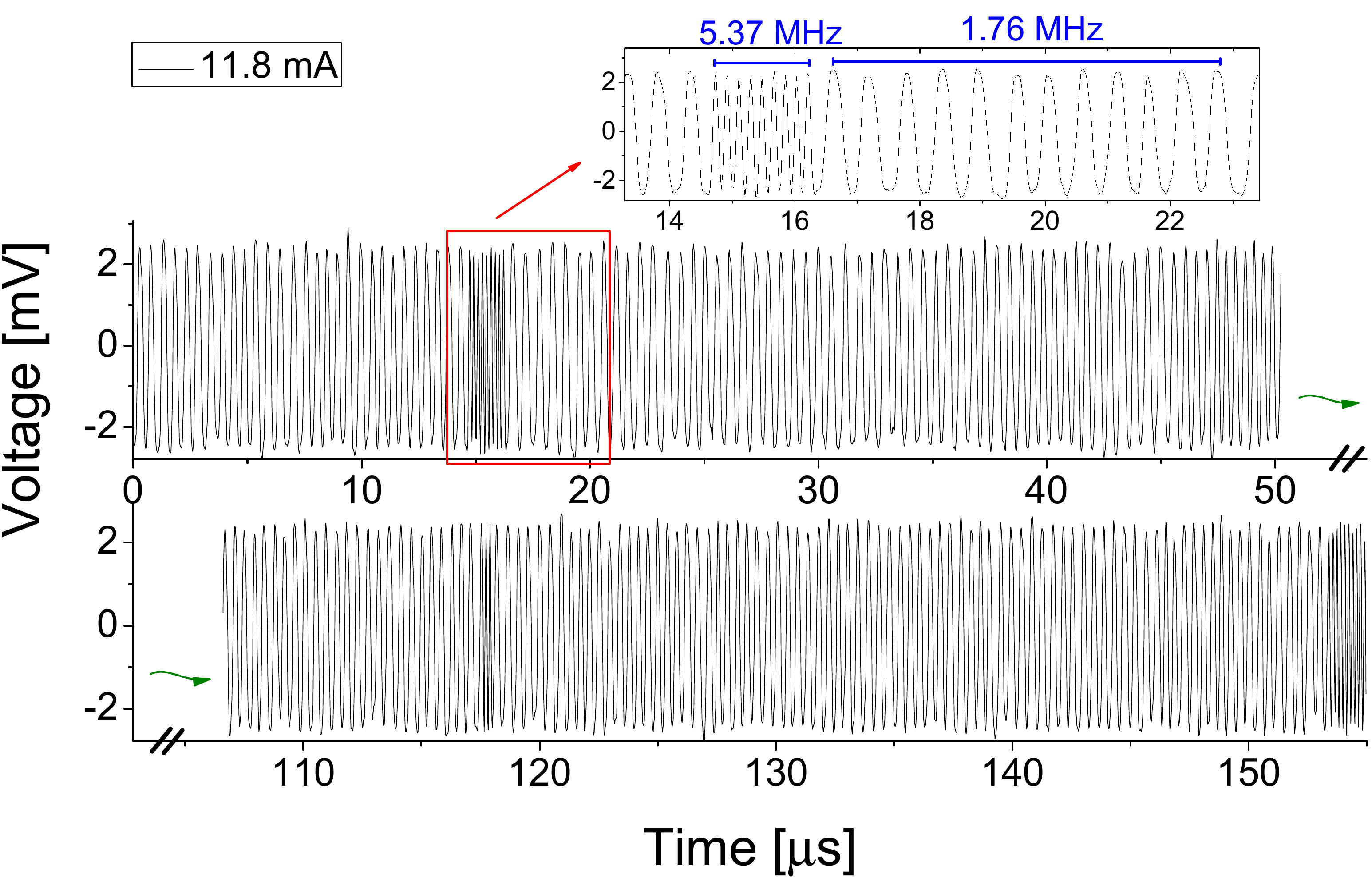}   } 
  \caption[Measured heterodyne voltage signal vs. time in the transition regime G$\rightarrow$C-state -- 2]{Measured heterodyne voltage signal vs. time in the transition regime at different current values, corresponding to the spectra shown in fig. \ref{fig_C-state:spectra_experiment_RTN-2}. The given frequencies (blue) are average values over the corresponding marked sequence.  
}
  \label{fig_C-state:TimeTraces_different_conditions}
\end{figure}

In fig. \ref{fig_C-state:TimeTraces-b} \& \ref{fig_C-state:TimeTraces-c}, we show time traces recorded corresponding to the spectra in fig. \ref{fig_C-state:spectra_experiment_RTN-2}. 
At lower currents in the transition interval of this measurement ($\sim [8.9;10.5]\,$mA), we observe voltage signals as shown in fig. \ref{fig_C-state:TimeTraces-b}.
 Here, the oscillation frequency is broader dispersed. 
 Characteristically, we observe kinks in the signal as e.g. at $5$ or at $5.3\,\upmu$s and marked by green arrows in fig. \ref{fig_C-state:TimeTraces-b}, which is interpreted as a frequent state-change and/or pinning of the vortex core at the disk boundary. 
At higher currents in the transition interval ($\sim [10.6;12.5]\,$mA), we find voltage time traces as shown in fig. \ref{fig_C-state:TimeTraces-c}. 
A stable oscillation with a specific frequency over a long range of time ($\sim 100\,\upmu$s) is observed. 
Beyond this time period, the system changes its state to another frequency of $\sim 4\,$MHz difference for a short duration. 
In this interval, the system oscillates mainly in the C-state, but rarely changes to the G-state where it performs several periods. 
Pinning at the potential barrier in this regime does not play a significant role and the state change is rather discrete.

\subsection{Dependence of the mode transition on the magnetic field}

\begin{figure}[bth!]
  \centering  
  \subfloat[  \label{fig_C-state:transition_vs_H-a}]
  {  
  \includegraphics[width=0.4425\textwidth]{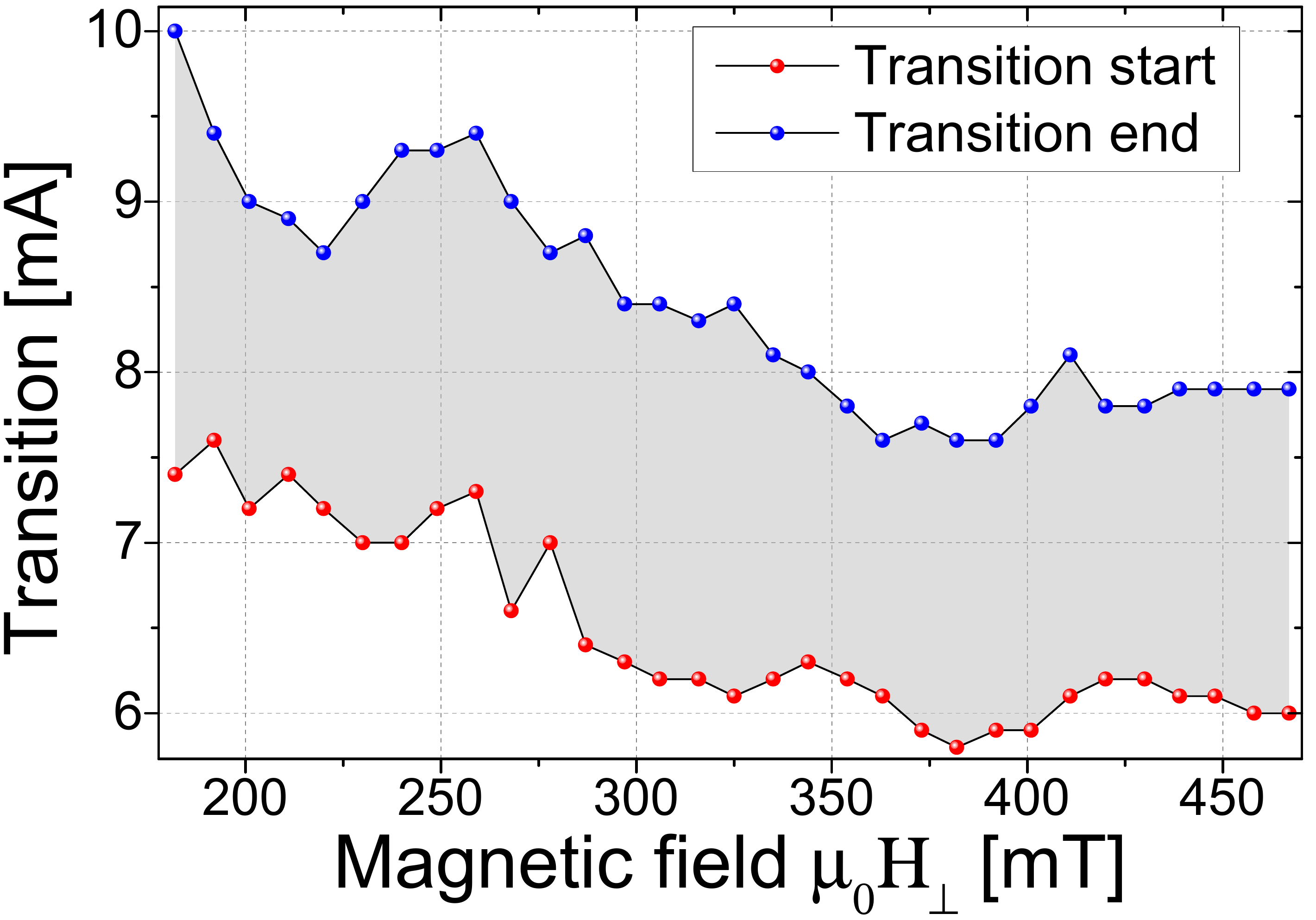}  }   
  \hfill 
 \subfloat[  \label{fig_C-state:f_vs_H-b}] 
  { 
  \includegraphics[width=0.436\columnwidth]{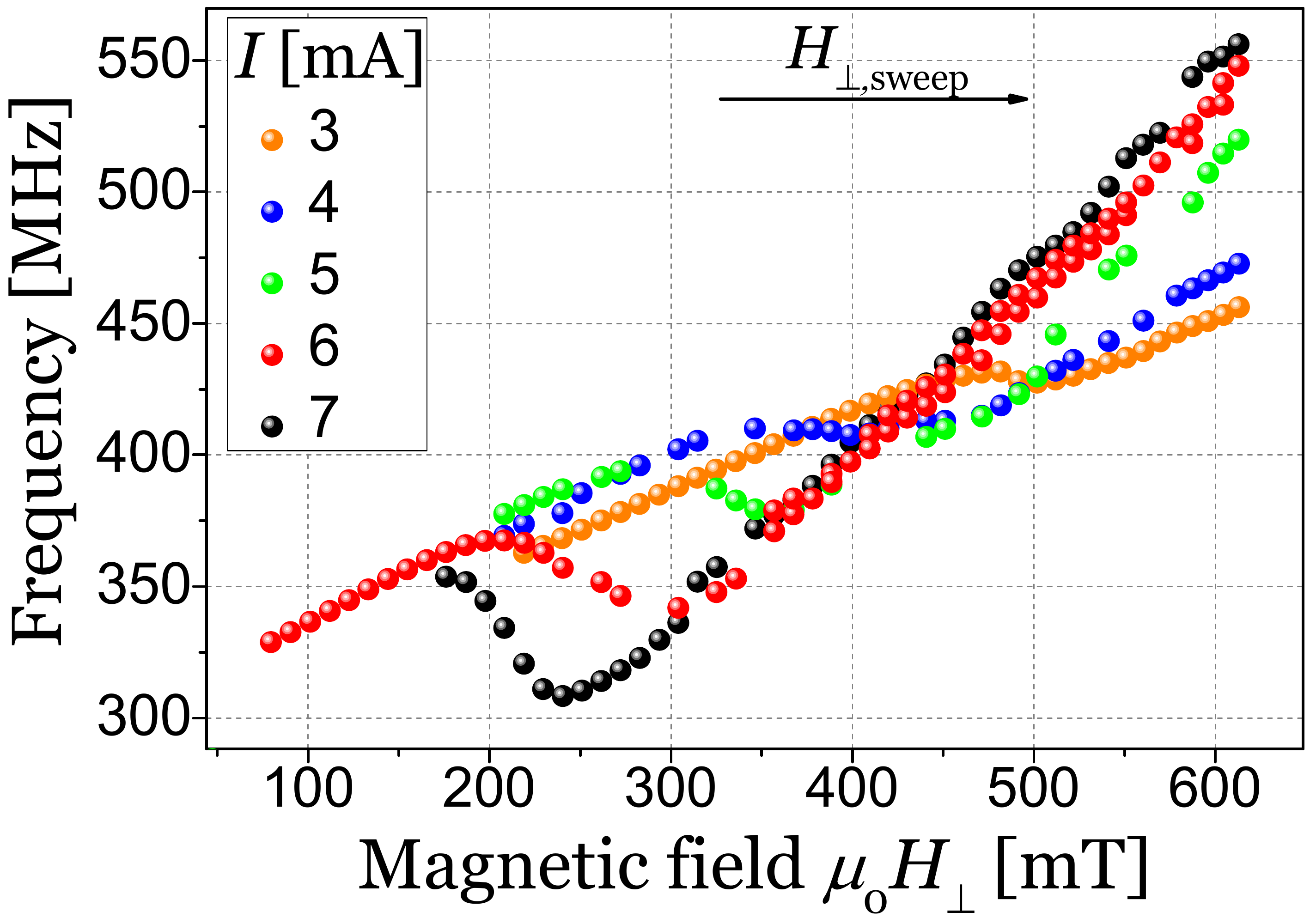}   } 
  \caption[State transition dependence on the magnetic field]{(a) Evolution of the transition range between G- and C-state with the  applied magnetic field $\mu_0 H_{\perp}$. (b) Frequency vs. $\mu_0 H_{\perp}$ at constant dc current values (different sample). 
  }
  \label{fig_C-state:dependence_vs_H}
\end{figure}

In this section, we investigate the dependence of the mode transition on the applied out-of-plane magnetic field $H_{\perp}$. 
Regarding fig. \ref{fig_C-state:transition_vs_H-a}, it is observed that an applied field $H_{\perp}$ lowers the energy barrier and thus facilitates the described transition between G- and C-state.  
We attribute this fact to the increase of the strength of the driving spin torque force acting on the vortex core, which is determined by the out-of-plane component of the STVO's fixed layer which grows with  $H_{\perp}$  \cite{Dussaux2012}. 
Furthermore, the applied magnetic field deforms the magnetic vortex distribution (see Ref. \cite{Dussaux2012}). Since the magnetization $z$-component at the boundary increases\cite{Dussaux2012}, the transition energy barrier is additionally lowered. 
Note that the results shown in fig. \ref{fig_C-state:dependence_vs_H} are measured for a different sample than the ones shown beforehand, qualitatively showing the same characteristics but with different values.


In fig. \ref{fig_C-state:f_vs_H-b}, we plot the frequency evolution vs. $\mu_0 H_{\perp}$ and observe that the different modes can also be triggered by the magnetic field. 
The transition between the states occurs for lower fields if $I_{dc}$ is higher, in agreement with the above interpretation of a higher spin torque strength facilitating the transition. 
Note that the slope $df/dH_{\perp}$ in the C- is larger than in the G-state, which might be due to a changed influence of the current induced Oersted field on the dynamics.

%

\section{Simulation}

The micromagnetic simulations are performed with mumax3 \cite{mumax}. 
The dimension of the STVO's free layer (nanodisk diameter $300\,$nm, $7\,$nm thick NiFe free layer) is discretized into $86\times 86\times 1$ cells. A NiFe saturation magnetization of $8\cdot 10^5\,$A/m, exchange stiffness $A_{ex}=1.3\cdot 10^{-11}\,$J/m, and damping $\alpha=0.01$ are chosen. 
In order to solve the Landau-Lifshitz-Gilbert-Slonczewski equation of magnetization dynamics, a standard Runge-Kutta 4/5 ODE solver has been used. 
Temperature is either set to $T=300\,$K or $T=0\,$K.

  \begin{wrapfigure}[17]{r}{0.467\columnwidth}
\centering
\vspace{-0.68cm}
\includegraphics[width=0.906\columnwidth]{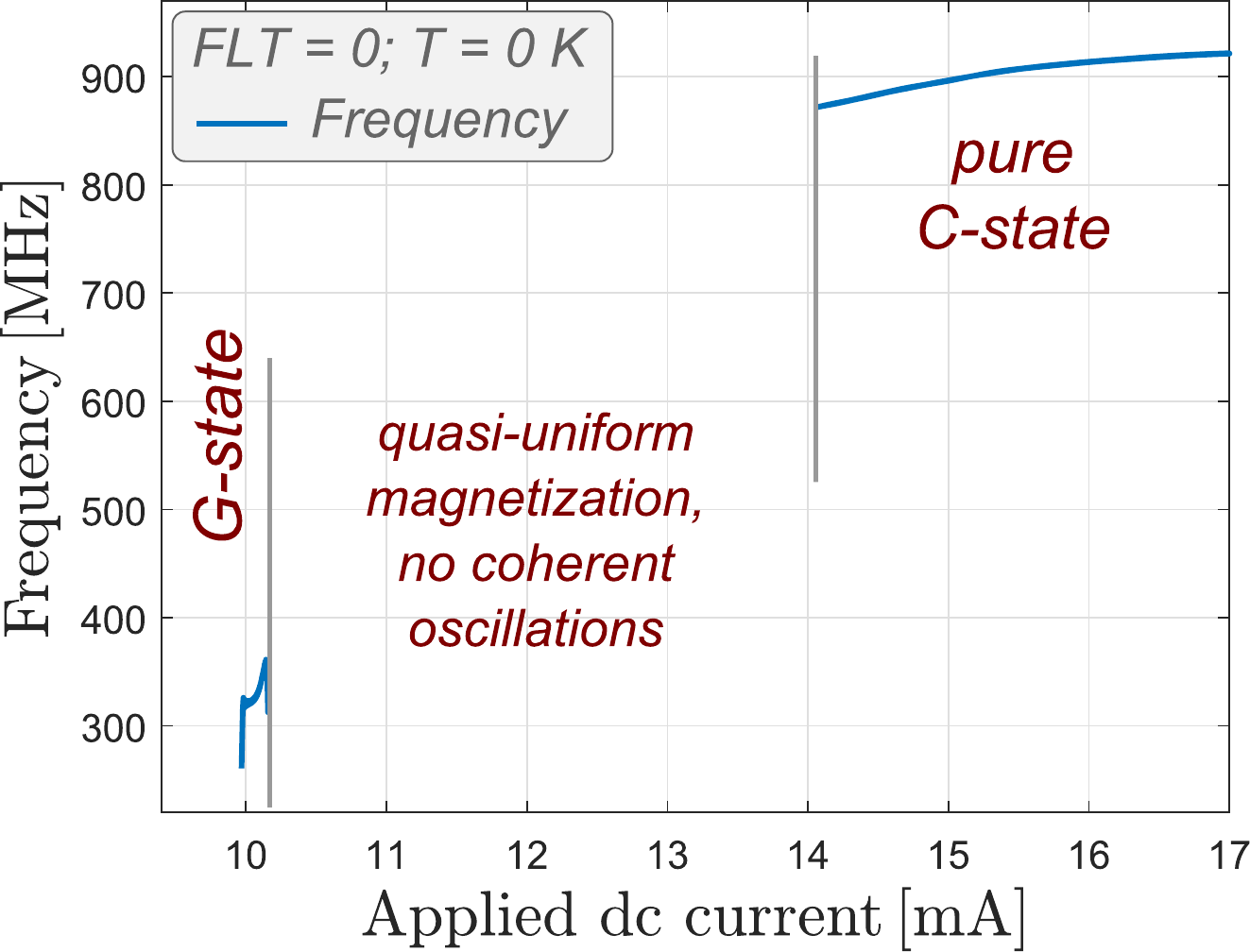}
\caption[Micromagnetic simulation of G- and C-state without field-like torque and thermal noise]{Micromagnetic simulation without field-like torque and thermal noise: Frequency evolution and magnetic states as a function of the applied current $I_{dc}$.
 }
\label{fig_C-state:sim_T=0_FLT=0}
\end{wrapfigure}

If the field-like torque is set to zero, the vortex motion does not exhibit an asymmetric component any more. 
The situation is shown in fig. \ref{fig_C-state:sim_T=0_FLT=0} for $T=0\,$K: In a small range at lower currents $I_{dc} \in \sim  [9.9; 10.2]\,$mA, above the current threshold for auto-oscillations, the gyrotropic motion is observed until at $I_{dc} \approx 10.2\,$mA, the vortex core is expulsed. 
The free layer is then quasi uniformly magnetized between $I_{dc} \approx 10.2\,$mA and $I_{dc} \approx 14\,$mA without considerable dynamics. At $I_{dc} \approx 14\,$mA, the C-state oscillations start, as shown in fig. \ref{fig_C-state:sim_T=0_FLT=0}. 
Note that here, if $T\neq 0\,$K in the presence of thermal noise, the transition between the G- and the C-state would be more continuous  with a smaller range without oscillations.  
However, the C-state in fig. \ref{fig_C-state:sim_T=0_FLT=0} has a significantly increased frequency of $\sim 900\,$MHz compared to the G-state and can be seen as a \textit{pure} C-state where no renucleation of the vortex core inside the nanodisk takes place. 
This observation is in agreement with findings in Ref. \cite{Kawada2014}, where similar measurements were conducted in GMR-based STVOs (i.e. negligible field-like torque) and a highly frequency increased C-state dynamic mode in the GHz range has been observed. 
The results highlight the role of the field-like torque, present in TMR based STVOs, to induce vortex C-state dynamics beyond the typical gyrotropic motion.

\bibliography{literatur_promo}